\documentclass[11pt]{article}

\usepackage{graphicx}
\usepackage[byname]{smartref}
\usepackage[authoryear]{natbib}
\usepackage{amsmath}
\usepackage{subfigure}
\usepackage{enumerate}
\usepackage{color}
\usepackage{authblk}
\usepackage[margin=2.5cm]{geometry}

\begin{document}

\title{Experiments on the vortex wake of a swimming knifefish}
\date{}

\author[1,2]{Zachary J. Taylor}
\author[1]{Alexander Liberzon}
\author[3]{Roi Gurka}
\author[4]{Roi Holzman}
\author[2]{Thomas Reesbeck}
\author[2]{F. Javier Diez\thanks{diez@jove.rutgers.edu}}

\affil[1]{School of Mechanical Engineering, Tel Aviv University, Tel Aviv, Israel}
\affil[2]{Department of Mechanical and Aerospace Engineering, Rutgers, The State University of New Jersey, Piscataway, NJ, USA}
\affil[3]{Department of Mechanical Engineering, Ben-Gurion University, Beer-Sheva, Israel}
\affil[4]{Department of Zoology, Tel Aviv University and the Inter-University Institute for Marine Sciences in Eilat, Eilat, Israel}    

\renewcommand\Authands{ and }

\maketitle

\begin{abstract} The knifefish species propels itself by generating a reverse K\'arm\'an street from an anal fin and without significantly moving its body. This unique feature makes this species' propulsion method highly efficient \citep{Blake1983}. It has been suggested that there is an optimal swimming range for fish based on the amplitude and frequency of the reverse K\'arm\'an street. Experiments have been performed to measure the ratio between the amplitude and wavelength of vortices in the wake of a knifefish. It is suggested that by optimizing the thrust created by the reverse K\'arm\'an street the wave efficiency can be estimated for a given spacing ratio, and present observations have an average value of 0.89. The relationship established between spacing ratio and wave efficiency, in addition to the measured parameters, will be invaluable for bio-inspired designs based on the knifefish.
\end{abstract}

\section{Introduction}
\label{intro}

In addition to understanding the ecological and evolutionary forces that shape animal diversity, the knowledge of fish swimming mechanisms is becoming increasingly important for the design and implementation of underwater vehicles.  It is natural that designers look to fish for design inspiration since fish swimming has been exposed to evolutionary forces that could select for optimal swimming \citep[e.g.,][]{Sfakiotakis1999}.  Fish inhabit a wide array of habitats which place different performance demands. Thus, different fish species evolved swimming behaviours suitable for different habitat and task, and there exists a wide range of propulsive methods \citep{Sfakiotakis1999}.  One of the swimming types recently suggested for use in underwater vehicles is that known as gymnotiform motion \citep{Blake1983}.  For their swimming robot, \citet{Shirgaonkar2008} have taken inspiration from one of the gymnotiform swimmers: the knifefish.  In the current study the focus is on experiments of a live knifefish species (shown in Fig. \ref{fig:knifefish}). 

\par The characteristic feature of gymnotiform motion is the anal fin (Fig. \ref{fig:knifefish}) that can be used for thrust by creating a travelling sinusoidal wave along the length of its body \citep{Blake1983}.  Thus, unlike other forms of fish locomotion, the body can be kept relatively rigid while swimming.  While this form of swimming does not produce the greatest thrust, it is known to have high hydromechanical efficiency ranging upwards of 80\% \citep{Blake1983}.  In addition, the direction of the wave can be reversed implying that the fish can swim backwards as easily as forwards.  Knifefish hunt for their prey by detecting disturbances in an electric field which suggests one of the reasons for this swimming behavior -- the prey is detected by swimming past it and a quick change of direction ensures that it does not escape after being detected \citep{Shirgaonkar2008}. According to \citet{Blake1983} it is unclear whether the electric sense preceded the gymnotiform motion in the evolution of the species. However, similar swimming behaviours are also observed in other weakly electric fish species, suggesting that this swimming mode co-evolved with electrical sense. 

\par The vortex wake behind fish resembles a K\'{a}rm\'{a}n vortex street -- only in reverse \citep{Triantafyllou1991}. Through experiments using a flapping foil, \citet{Triantafyllou1991} suggested that there is a narrow range of non-dimensional tail beating frequencies for optimal thrust between $0.2<\mathrm{St}<0.4$.  To calculate thrust, \citet{Streitlien1998} have reviewed different methods based on measurements of the wake and/or body of the fish.  One of the three models they used was K\'{a}rm\'{a}n's formulation for the drag induced by a vortex street with the suitable modification of reversing the sign of the circulation.  Even though this model is relatively simplistic, it was shown to be accurate to within 10\% \citep{Streitlien1998}.  One of the challenges of using the K\'{a}rm\'{a}n model experimentally is the apparent need to measure the circulation; however, the thrust of the vortex street need not depend on the circulation directly since the speed of the vortex system, $U_s$, and the circulation are linked \citep[e.g.,][]{Weihs1972}.  If the vortex street thrust coefficient is defined as $C_{T_s}=T_s/\frac{1}{2}\rho U_f^2 a$ where $T_s$ is the thrust; $\rho$ the density of water; $U_f$ the speed of the fish; and $a$ the vortex wavelength then the dependency of $C_{T_s}$ on the spacing ratio $b/a$ can be written similarly to \citet{Bearman1967}  
\begin{equation}
\resizebox{0.95\width}{!}
{$C_{T_s}=\frac{4}{\pi}\left(\frac{U_s}{U_f}\right)^2\left[\left(\frac{U_f}{U_s}-2\right)\frac{\pi b}{a}\mathrm{coth}\frac{\pi b}{a}+\mathrm{coth}^2\frac{\pi b}{a}\right]$.}
\label{eq:thrust} 
\end{equation} 
\par The goal of the current study is to use experimental data to quantify the range of spacing ratios that could be expected in the wake of knifefish.  These measurements allow the approximation of knifefish swimming efficiency, and can help establish design criteria for those using the knifefish species in their bio-inspired designs \citep[e.g.,][]{Curet2011}.   

\section{Details of the experiments}
\label{sec:experiments}

\begin{figure} \centering
\includegraphics[width=4.5cm]{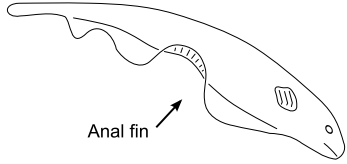} 
\caption{Sketch of a knifefish with the unique anal fin identified.} 
\label{fig:knifefish} 
\end{figure}

A sketch of the fish is shown in Fig. \ref{fig:knifefish} identifying the unique anal fin.  In order to adhere to appropriate policies, regulations and laws for handling and care of vertebrate animals, the tests followed protocol 08-017 approved by Rutgers University Animal Care and Facilities Committee for studying the African Knifefish (Xenomystus Nigri) swimming methods. Following the protocol, all tests were performed only by approved personnel at the Rutgers facility. Tests were performed in a recirculating tank constructed out of acrylic that allowed holding the fish in the laser field with the possibility of a flow field to stimulate swimming. A 373 W water pump generated flow speeds up to 11 cm/s in the test section with dimensions $8 \times 8 \times 38$ cm$^3$. 
\par The flow data were taken underneath the fish in a horizontal plane using two-component Particle Image Velocimetry (PIV).  The PIV system comprised a Nd:YAG laser (New Wave Research Solo PIV 120) and a digital PIV CCD Camera (TSI Powerview Plus 2 MP) at 14.5 Hz.  In the current study the focus is on six instances when the wake of the knifefish was in the field-of-view ($11\times 8$ cm$^2$) of the camera: three while the fish was swimming forwards (Cases 1-3) and three while the fish was swimming backwards (Cases 4-6).  It should be noted that although the wake was in the field-of-view of the cameras there were only a few snapshots where a sufficient portion of the fish appeared in the image to determine its speed. 

\section{Results and Discussion}
\label{sec:results}

\subsection{Vortex identification}
\label{sec:vortexID}

To measure the spacing ratio in the wake of the knifefish it is necessary to identify the vortex centers.  Although there is no formal definition of a vortex, locations in the flow where local pathlines form circular patterns are identified by computing the so-called `swirling strength', $\lambda_{ci}$.  The swirling strength is defined as the complex portion of the complex pair of eigenvalues of the two-dimensional velocity gradient tensor (if the eigenvalues are real then there is no vortex), and this method is often used on two-dimensional PIV data \citep[e.g.,][]{Adrian2000}.  Once the value of $\lambda_{ci}$ is computed for each point in a velocity map, the locations with the highest swirling strength are identified as the vortex centers.   

\par
\begin{figure*}[ht] 
\centering
\subfigure[]{\label{fig:vortexID-a}\includegraphics[width=0.4\textwidth]{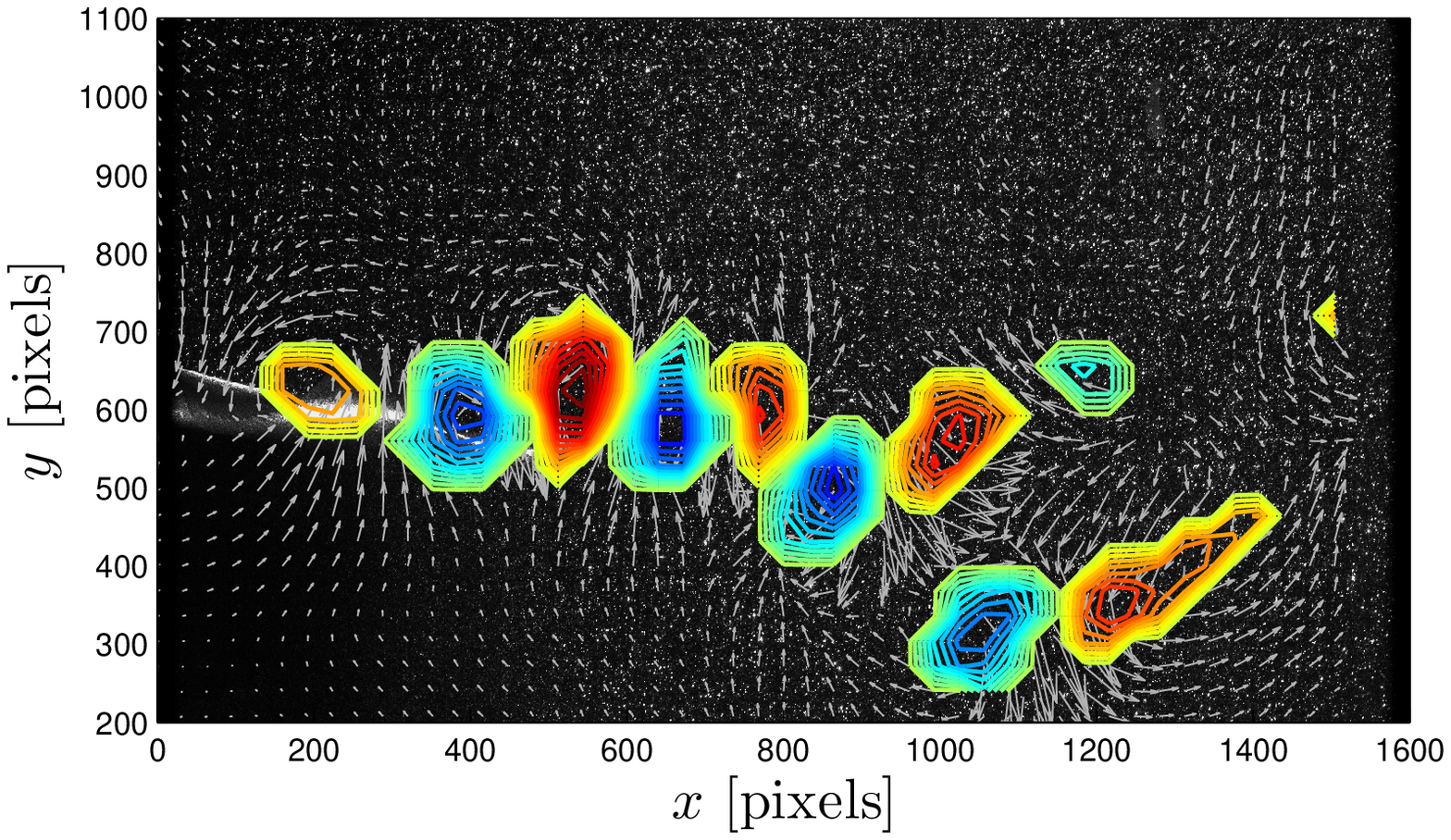}}
\subfigure[]{\label{fig:vortexID-b}\includegraphics[width=0.4\textwidth]{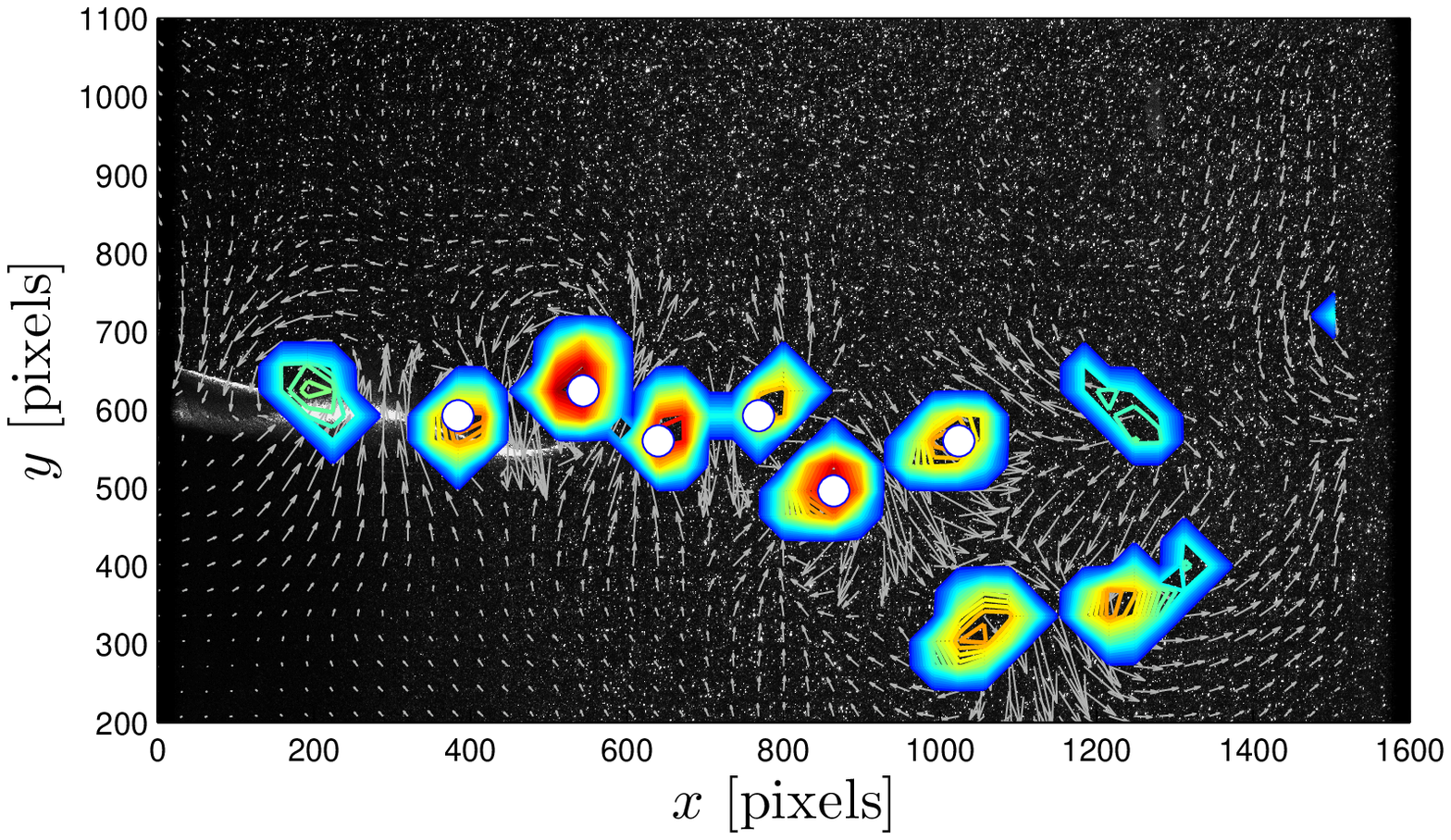}}
\caption{An example of an instantaneous PIV image with velocity vectors overlaid in the laboratory frame-of-reference.  In \subref{fig:vortexID-a} the contours represent the vorticity field, and in \subref{fig:vortexID-b} are contours of the swirling strength, $\lambda_{ci}$.} 
\label{fig:vortexID}
\end{figure*}

 A sample velocity map has been plotted on top of a PIV image with vorticity contours in Fig. \ref{fig:vortexID-a} and swirling strength contours in Fig. \ref{fig:vortexID-b}.  Also shown in Fig. \ref{fig:vortexID-b} are the identified locations of high swirling strength highlighting the vortex centers.   

\subsection{Vortex Spacing} \label{sec:vortexChars}

\begin{figure} 
\centering
\includegraphics[width=6.5cm]{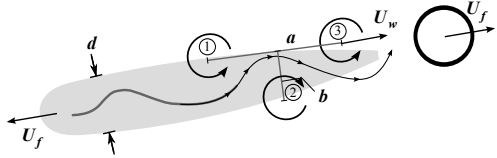} 
\caption{Left: the speed of the knifefish $U_f$; the speed of the vortices $U_w$;  the width of the body $d$; the numbering of the vortices; the vortex wavelength $a$; and cross-stream distance between the vortex rows $b$.  Right: the analogous circular cylinder moving with the same velocity in the opposite direction.} 
\label{fig:vortexspacing} 
\end{figure}

One of the classical descriptions of vortex street wakes is the ratio of wavelength, $a$, and the cross-stream distance between the two rows of vortices, $b$ \citep{Bearman1967}.  These lengths have been computed according to Fig. \ref{fig:vortexspacing}.  The vortex wavelength is computed as the distance between vortices 1 and 3 and the cross-stream distance, $b$, is defined as the shortest distance between the vortex at $(x_2,y_2)$ and the line connecting vortices at $(x_1,y_1)$ and $(x_3,y_3)$. 
\par The number of vortex triplets used to estimate the spacing ratio ranged between 1-4 in each PIV frame, and each frame-average spacing ratio is presented in Table \ref{tab:parameters}.  A significant range of spacing ratios is observed in Table \ref{tab:parameters} which agrees with the variety of spacing ratios reported in the literature \citep[e.g.,][]{Bearman1967}.  One of the particular features of the knifefish is that the vortices are created by the sinusoidal shape and movement of the fin rather than the beating of its tail.  Thus, rather than exclusively creating vortices behind its body, the knifefish creates vortices underneath its body (e.g., Fig. \ref{fig:vortexID}).  This swimming means it can instantaneously change the spacing ratio of the vortices since it is governed by the shape of the anal fin.  For fish and robotic fish, \citet{Epps2009} reported a wide range of spacing ratios (0.2-0.8) which range noticeably higher than those found for the knifefish. 
\begin{table}
\caption{Wake parameters measured for different cases.  Each line is the average value for each PIV frame.  The table is divided into forward (left side) and backward (right side) swimming.}
\label{tab:parameters} 
\begin{tabular}{llcl|llcl}
\hline\noalign{\smallskip}
No. & $b/a$ & $U_f/U_w$  & St & No. & $b/a$ & $U_f/U_w$  & St\\
\noalign{\smallskip}\hline\noalign{\smallskip}
1	&	0.37	&	0.79	&	0.29	&	4	&	0.26	&	0.90	&	0.23	\\
	&	0.29	&	0.87	&	0.25	&		&	0.19	&	0.97	&	0.18	\\
	&	0.27	&	0.89	&	0.24	&		&	0.04	&	1.00	&	0.04	\\
2	&	0.35	&	0.81	&	0.28	&		&	0.13	&	0.99	&	0.13	\\
	&	0.30	&	0.86	&	0.26	&		&	0.13	&	0.99	&	0.13	\\
	&	0.50	&	0.72	&	0.36	&		&	0.08	&	1.00	&	0.08	\\
	&	0.22	&	0.94	&	0.21	&	5	&	0.17	&	0.98	&	0.17	\\
	&	0.44	&	0.74	&	0.33	&		&	0.32	&	0.84	&	0.27	\\
	&	0.25	&	0.91	&	0.23	&	6	&	0.14	&	0.99	&	0.14	\\
	&	0.19	&	0.96	&	0.18	\\								
	&	0.26	&	0.91	&	0.23	\\								
3	&	0.47	&	0.73	&	0.34	\\								
	&	0.26	&	0.91	&	0.23	\\								
	&	0.41	&	0.76	&	0.31	\\								
	&	0.16	&	0.98	&	0.16	\\								
	&	0.24	&	0.92	&	0.22	\\								
\noalign{\smallskip}\hline
\end{tabular}
\end{table}

\subsection{Efficiency}
\label{sec:efficiency}

The research on vortex street wakes by \citet{Bearman1967} used K\'{a}rm\'{a}n's drag equation for a vortex street and the $\partial C_{Ds}/\partial (b/a)=0$ stability criterion proposed by Kronauer to link the spacing ratio and the convection speed of vortices in the wake of bluff bodies.  The schematic shown in Fig. \ref{fig:vortexspacing} shows the reverse K\'{a}rm\'{a}n street generated in the wake of the knifefish and the corresponding motion in the opposite direction of a circular cylinder.  Thus, if the fish generates the same vortex arrangement as the cylinder while moving in the opposite direction, the drag induced by the vortex street is analogous to the thrust produced by the fish (Eq. \ref{eq:thrust}).  Moreover, re-framing the Kronauer stability criterion to maximize thrust (instead of minimizing drag) based on the vortex spacing then the speed of the vortex system, $U_s$, can be related to the speed of the fish, $U_f$, similarly to \citet{Bearman1967}.  Differentiating Eq. \ref{eq:thrust} with respect to the spacing ratio $b/a$ and setting it to zero yields Eq. \ref{eq:ratio} where $k=\pi b/a$ is substituted for compactness. 
\begin{equation} 
\frac{U_f}{U_s}=2\left(1+\frac{1}{\mathrm{sinh}^2k - k\mathrm{tanh}k}\right) 
\label{eq:ratio} 
\end{equation} 
One of the classical measures of swimming efficiency is the so-called `wave slip', or wave efficiency, defined as the ratio $U_f/U_w$ where $U_w$ is the wave speed, or convection speed, of the vortex street \citep{Sfakiotakis1999}.  In the assumptions behind Eq. \ref{eq:thrust}, the wave speed of the vortices is defined as $U_w=U_f+U_s$.  Thus, the wave efficiency is calculated using Eq. \ref{eq:ratio} and 
\begin{equation} 
\frac{U_f}{U_w}=\frac{1}{1+U_s/U_f}.
\label{eq:slip} 
\end{equation} 
These data are added to Table \ref{tab:parameters} for each PIV measurement in all 6 cases.  The wave efficiency is not the same as the hydromechanical efficiency cited previously by \citet{Blake1983}.  However, the range of wave efficiencies found here are roughly consistent with those reported by \citet{Curet2011} of 0.6-0.9.  The efficiencies approach unity as the spacing ratio approaches zero.  However, it is not possible for a fish to continue swimming if $b/a \to 0$.  As previously mentioned, since the fish did not often appear in the image at the same time as its wake, it was not possible to estimate $U_f$ very often from the current dataset.  However, when it was possible, the wave slip was computed using the convection speed of the vortices as the wave speed (taken as the speed at the center of the vortex).  The measured wave efficiency ranged between 0.85-0.89 which agrees with the predicted values based on the spacing ratios in Table \ref{tab:parameters}.   

\par In addition to the wave slip, it is well known that there is an optimal Strouhal number ($\mathrm{St}=fb/U_f$) range of $0.2 < \mathrm{St} < 0.4$ \citep{Triantafyllou1991} for propulsive wakes.  Since the frequency of the vortex street is $f=U_w/a$ the Strouhal number can be written as 
\begin{equation} 
\mathrm{St}=\frac{U_w}{U_f}\frac{b}{a}. 
\label{eq:St}
\end{equation} 
The Strouhal numbers have been added to Table \ref{tab:parameters} using the measured spacing ratios and Eq. \ref{eq:slip}, and for most cases the calculated Strouhal numbers fall inside the optimal range predicted by \citet{Triantafyllou1991}.  The cases falling outside of this optimal range correspond to cases when $b/a \to 0$.

\section{Conclusions}
\label{sec:conclusions}

Through PIV experiments on a swimming knifefish the spacing ratio of the wake is calculated for several repeated cases.  The PIV measurements also show that the knifefish can create a thrust-generating vortex street wake while a portion of that wake remains underneath its body.  This ability is due to the unique anal fin and one of the reasons for its superior swimming efficiency.  A link has been suggested between spacing ratio and efficiency (Eq. \ref{eq:slip}), thus it is not surprising that the spacing ratios measured for the knifefish are slightly smaller than in previous studies \citep{Epps2009} since knifefish are known to be among the most efficient swimmers \citep{Blake1983}.  While the two-dimensional approximations seem to match previously published wave efficiencies and optimal Strouhal numbers, \citet{Shirgaonkar2008} have shown the importance of the three-dimensional connections between the vortices and further studies are required to address how the neglected three-dimensionality affects the swimming efficiency of knifefish.   
\section*{Acknowledgments}
Z.J. Taylor is grateful to the Azrieli Foundation for the award of an Azrieli Fellowship.

%
\bibliographystyle{basic}      

\end{document}